# An Industry-Academia Partnership for Advancing Quantum Frontiers: Perspective from the U.S. Center for Quantum Technologies

**David Stewart[1,†], Gerardo Ortiz[2,‡], Peter M. Kogge[3,§], Ricardo S. Decca[4,#], Tongcang Li[1,5,6,*]**

[1]Purdue Quantum Science and Engineering Institute, Purdue University, West Lafayette, IN, USA
[2]Department of Physics, Indiana University, Bloomington, IN, USA
[3]Department of Computer Science and Engineering, University of Notre Dame, Notre Dame, IN, USA
[4]Department of Physics, Indiana University Indianapolis, Indianapolis, IN, USA
[5]Department of Physics and Astronomy, Purdue University, West Lafayette, IN, USA
[6]Elmore Family School of Electrical and Computer Engineering, Purdue University, West Lafayette, IN, USA
[†] davidstewart@purdue.edu
[‡] ortizg@iu.edu
[§] kogge@nd.edu
[#] rdecca@iu.edu
[*] tcli@purdue.edu

**ABSTRACT**

The U.S. Center for Quantum Technologies (CQT) is a multi-university consortium established under the National Science Foundation's (NSF) Industry-University Cooperative Research Centers (IUCRC) program. Led jointly by Purdue University, Indiana University (both Bloomington and Indianapolis campuses), and the University of Notre Dame, CQT integrates academic research with industrial and governmental collaboration to accelerate quantum innovation. This perspective outlines the consortium's strategic mission, interdisciplinary research agenda, and its role in shaping the future of quantum-enabled technologies through collaborative development, translational impact, and workforce cultivation.

**INTRODUCTION**

Quantum technologies are rapidly transforming the landscape of computation, communication, sensing, and security. The second quantum revolution—centered on the manipulation of quantum states for practical applications—requires coordinated efforts across academia, industry, and government. The U.S. Center for Quantum Technologies (CQT), established in 2022, exemplifies a collaborative model designed to meet this challenge. Operating under the IUCRC framework, CQT brings together three leading research universities in Indiana along with industry and government members to form a unified hub for quantum innovation.

The Center's mission is to develop quantum technologies that are scientifically rigorous, commercially viable, and societally impactful. By integrating foundational research with translational goals, CQT aims to address emerging technological challenges while fostering a robust quantum workforce.

**CONSORTIUM STRUCTURE AND STRATEGIC VISION**

CQT is jointly led by Purdue University, Indiana University (both Bloomington and Indianapolis campuses), and the University of Notre Dame. This structure leverages complementary strengths in quantum science, engineering, and computing across institutions. The IUCRC model facilitates shared governance, with industry and government partners actively participating in research prioritization, mentoring, and progress review. The Center currently has 12 industry and government members— Air Force Research Lab, Amazon Web Services, Cummins, D-Wave, Eli Lilly, Entanglement Inc., Hewlett Packard Enterprise, Infleqtion, L3Harris, Peraton, Quantum Computing Inc., and Quantum Corridor—and is actively expanding its membership, inviting new partners to join in shaping the future of quantum technologies.

The strategic vision of CQT includes advancing interdisciplinary quantum research, enabling technology transfer through industry and government engagement, supporting scalable innovation in quantum computing, sensing and communications, and cultivating a quantum-ready workforce. The consortium's collaborative ecosystem amplifies the impact of individual institutions and accelerates innovation across Indiana and the country.

**RESEARCH THEMES AND SCIENTIFIC CONTRIBUTIONS**

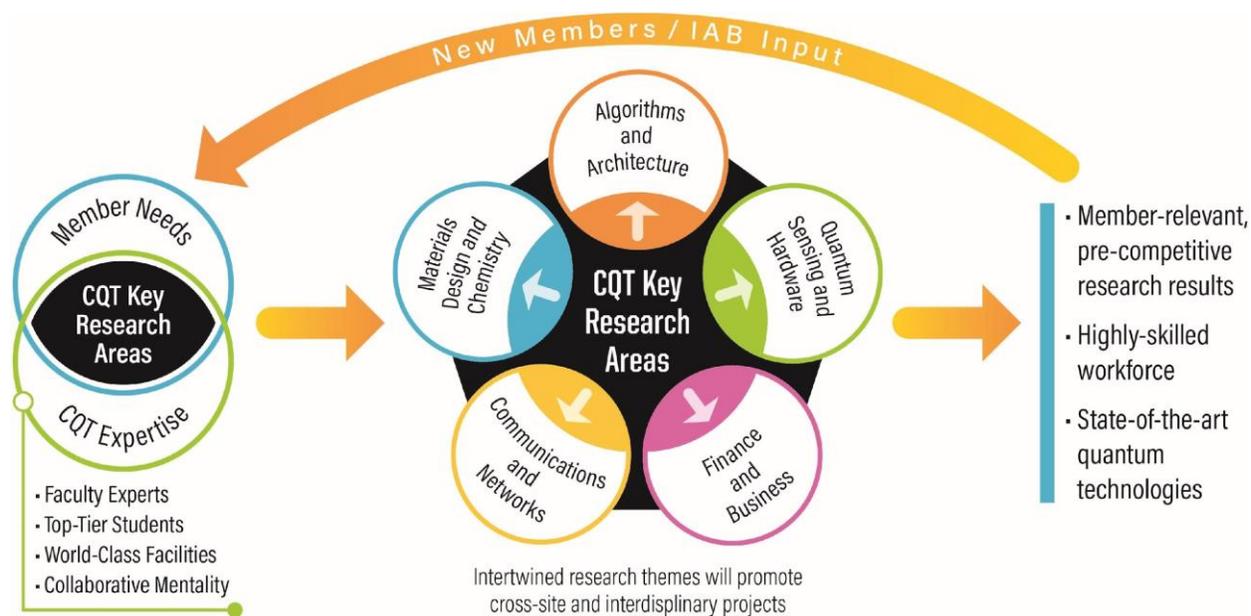

**Figure 1**. Center for Quantum Technologies (CQT) Key Research Areas highlighting interdisciplinary projects driven by member needs and university expertise to produce state-of-the-art quantum technologies and a highly-skilled workforce.

CQT's research portfolio spans a diverse array of quantum domains with a focus on algorithms and architecture, materials design and chemistry, quantum sensing and hardware, finance and business, and communications and networks (Figure 1). The Center has already produced significant research results over the first two years of operation.

Quantum computing research at CQT has focused on algorithm development, benchmarking, and architecture innovation. By mapping classical natural language processing (NLP) embeddings used in state-of-the-art Generative AI systems (e.g., OpenAI's GPT 4 and Anthropic's Claude 4) into compact quantum states, researchers have demonstrated almost lossless integration of classically trained AI models. In downstream tasks, for example lexical or text similarity computation, or text classification, the classical-to-quantum mapping approaches were tested on simulators and the IBM quantum computer ibm_kyoto [1,2,3]. In parallel, hybrid quantum–classical algorithms have been developed to solve large graph optimization problems on Noisy Intermediate-Scale Quantum (NISQ) devices such as D-Wave quantum annealers and QuEra neutral-atom quantum computers [4]. Further, studies of the semantics of the crossover between the quantum and classical regimes are providing insight into improving how we express computations for such hybrid regimes. Finally, quantum computing techniques such as using QUBO and quantum annealing, are finding use in locating vulnerabilities in software due to malicious object injection.

In quantum materials, CQT researchers have combined quantum computers with advanced experimental techniques to uncover new behaviors. For example, a D-Wave quantum annealer was used in an unconventional coherent quantum quench mode to simulate the dynamics of geometrically frustrated two-dimensional XY models, revealing deviations from the quantum Kibble–Zurek scaling and the discovery of a quantum coarsening dynamics, demonstrating the power of quantum simulators for problems beyond classical reach [5]. Other work used neutron scattering and simulations to explain how defects stabilize unusual magnetic states in a frustrated magnet [6], while thin films of cobalt valence tautomer molecules were shown to switch conductivity with light, suggesting opportunities for light-controlled electronics [7]. In addition, CQT research into using solid neon and argon to trap single electrons and ions as ultraclean qubit platforms holds the promise of greatly improved qubit characteristics [8]. Furthermore, single spin defects in hexagonal boron nitride (hBN) have been created through ion implantation and used to detect individual $^{13}$C nuclear spins, demonstrating the potential of 2D materials for high-precision quantum sensing [9].

CQT researchers have also developed new photonic platforms to generate quantum states of light with applications in imaging, sensing, and communication. Researchers demonstrated the creation of superradiant and subradiant entangled states in pairs of organic molecules embedded in nanocrystals, tuned into resonance by laser illumination, providing a scalable route to collective quantum phenomena in subwavelength arrays [10]. Further work is engineering scalable and tunable topological insulator (TI) quantum dots to manifest emission and absorption in the far- and mid-infrared bands.

Further advances underscore the promise of engineered photonic nanostructures that boost light–matter interactions, paving the way for next-generation quantum optical devices. CMOS-compatible cuprous oxide films and two-dimensional Rydberg exciton arrays were realized as scalable platforms for strong photon–photon interactions [11,12]. Resonant nanophotonic structures were shown to reduce the effective dimensionality of quantum emitter ensembles [13], while driven-dissipative systems revealed spontaneous symmetry breaking, pattern formation in Rydberg polaritons [14], and robust time-crystal-like phases in a quantum optical system with Kerr nonlinearity [15]. Hybrid thermal atom–nanophotonic platforms were also developed, enabling tunable atomic interactions and strong nonlinearities for scalable quantum technologies [16].

Together, these research achievements illustrate CQT's ability to address critical challenges in quantum science, leveraging interdisciplinary expertise and strong academic–industry collaboration to advance both fundamental understanding and practical technologies.

**INDUSTRY AND GOVERNMENT ENGAGEMENT**

CQT's success is rooted in its deep engagement with industry and government stakeholders. Partners include major technology firms, defense contractors, pharmaceutical companies, and federal agencies, as noted earlier. These collaborators contribute funding, strategic direction, and domain expertise, while benefiting from early access to research outputs and talent pipelines. Industry and government members participate in semiannual meetings, project reviews, and student mentoring, creating a dynamic feedback loop between research and application. This multi-sector collaboration ensures that CQT's work remains relevant, scalable, and impactful.

In addition to early research and talent access, there are several other benefits for industry and government members, including leveraged research dollars, access to intellectual property, research risk mitigation, and cross-sector networking. CQT welcomes inquiries from organizations interested in membership.

**WORKFORCE DEVELOPMENT AND EDUCATIONAL MISSION**

A cornerstone of CQT's mission is the cultivation of a quantum-ready workforce. Students from all CQT institutions are embedded in interdisciplinary research teams, contributing to proposal development, experimental design, and dissemination of results. The Center's industry and government members provide direct mentorship to students over the course of the research, providing use-inspired training that better prepares them to make an immediate impact upon entering the workforce.

By integrating students into the full lifecycle of research and development, CQT fosters experiential learning and professional growth. The consortium's educational initiatives align with national priorities for quantum workforce development, ensuring a sustainable pipeline of skilled professionals.

CQT is also exploring ways to introduce younger audiences to quantum. Over two days in April of 2024, CQT along with the Purdue Quantum Science and Engineering Institute and the Quantum Science Center sent graduate students to Southport High School in Indianapolis, IN to introduce quantum technologies to high school students and share their excitement of pursuing a career in quantum science and technology.

**CONCLUSION**

The U.S. Center for Quantum Technologies represents a successful model for quantum research and innovation. Through its multi-university collaboration, strategic partnerships with industry and government, and commitment to education, CQT is advancing the frontiers of quantum science and engineering. As quantum technologies continue to evolve, the Center's integrated approach will play a critical role in shaping the future of science, technology, and society. To learn more about CQT, please visit: https://www.purdue.edu/cqt/. To learn more about the NSF IUCRC program, please visit: https://iucrc.nsf.gov/centers/center-for-quantum-technologies/.

**Acknowledgments**
This work was supported by the U.S. National Science Foundation through the Center for Quantum Technologies (CQT), a consortium of Purdue University, Indiana University, and the University of Notre Dame, under grant numbers 2224960, 2224928, and 2224985.

**REFERENCES**
1. Cavar D and Zhang C 2024 Semantic Similarities Using Classical Embeddings in Quantum NLP *IEEE International Conference on Quantum Computing and Engineering (QCE)* pp. 450-451 https://doi.org/10.1109/QCE60285.2024.10350
2. Zhang C et al. 2024 Entangled Meanings: Classification and Ambiguity Resolution in QNLP *IEEE International Conference on Quantum Computing and Engineering (QCE)* pp 97–102 https://doi.org/10.1109/QCE60285.2024.10260
3. Cavar D and Parukola K R 2025 Word and Text Similarity Using Classical Word Embeddings in Quantum NLP Systems *IEEE International Conference on Acoustics, Speech, and Signal Processing Workshops (ICASSPW)* pp. 1-5 https://doi.org/10.1109/ICASSPW65056.2025.11010999
4. Xu H, Pothen A 2024 Divide and Conquer-based Quantum Algorithms for Maximum Independent Set on Large Separable Graphs *IEEE International Conference on Quantum Computing and Engineering (QCE)* pp. 87-97 https://doi.org/10.1109/QCE60285.2024.00020
5. Ali A et al. 2024 Quantum quench dynamics of geometrically frustrated Ising models *Nat. Commun.* **15** 10756 https://doi.org/10.1038/s41467-024-54701-4
6. Khundzakishvili G et al. 2025 Criticality and Magnetic Phases of Ising Shastry–Sutherland Candidate Holmium Tetraboride *Mater.* **18** 2504 https://doi.org/10.3390/ma18112504
7. Phillips J P et al. 2024 Conductance fluctuations in cobalt valence tautomer molecular thin films *Dalton Trans.* **53** 17571 https://doi.org/10.1039/d4dt02213k
8. Li X et al. 2025 Noise-resilient solid host for electron qubits above 100 mK *arXiv:2502.01005* https://doi.org/10.48550/arXiv.2502.01005
9. Gao X et al. 2025 Single nuclear spin detection and control in a van der Waals material *Nature* **643**, 943 https://doi.org/10.1038/s41586-025-09258-7
10. Lange C M et al. 2025 Superradiant and subradiant states in lifetime-limited organic molecules through laser-induced tuning *Nat. Phys.* **21** 556–563 https://doi.org/10.1038/s41567-024-02404-4
11. Barua K et al. 2025 Bottom-up fabrication of 2D Rydberg exciton arrays in cuprous oxide *Commun. Mater.* **6** 21 https://doi.org/10.1038/s43246-025-00742-1
12. DeLange J et al. 2023 Highly-excited Rydberg excitons in synthetic thin-film cuprous oxide *Sci. Rep.* **13** 16881 https://doi.org/10.1038/s41598-023-41465-y
13. Boddeti A K et al. 2024 Reducing Effective System Dimensionality with Long-Range Collective Dipole-Dipole Interactions *Phys. Rev. Lett.* **132** 173803 https://doi.org/10.1103/PhysRevLett.132.173803
14. Alaeian H and Walther V 2024 Rise and fall of patterns in driven-dissipative Rydberg polaritons *Phys. Rev. Res.* **6** 033065 https://doi.org/10.1103/PhysRevResearch.6.033065
15. Alaeian H et al. 2024 Noise-resilient phase transitions and limit-cycles in coupled Kerr oscillators *New J. Phys.* **26** 023021 https://doi.org/10.1088/1367-2630/ad2414
16. Alaeian H et al. 2024 Manipulating the dipolar interactions and cooperative effects in confined geometries *New J. Phys.* **26** 055001 https://doi.org/10.1088/1367-2630/ad42c7